\documentclass[journal=jz,manuscript=letter]{achemso}
\usepackage[utf8]{inputenc}
\usepackage{amsmath, amssymb, amsthm}
\usepackage{graphicx}
\usepackage{soul}
\usepackage[usenames,dvipsnames]{xcolor}

\title{On the Road Again: How the $N$-Lane TASEP and Local Reservoirs Drive Jamming in Axonal Transport}

\author{Gustav Jennetten}
\affiliation{Department of Mathematics, Florida State University, Tallahassee, FL 32306, USA}

\author{Bhargav Karamched}
\affiliation{Department of Mathematics, Florida State University, Tallahassee, FL 32306, USA}
\alsoaffiliation{Institute of Molecular Biophysics, Florida State University, Tallahassee, FL 32306, USA}
\alsoaffiliation{Program in Neuroscience, Florida State University, Tallahassee, FL 32306, USA}
\email{bkaramched@fsu.edu}

\begin{document}

\begin{tocentry}
    \centering
    \includegraphics[width=3.25in, height=1.75in, keepaspectratio]{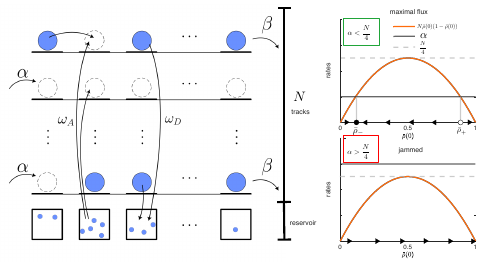}
\end{tocentry}

\begin{abstract}
    Intracellular transport by molecular motors along microtubules is essential for cellular function, particularly in neuronal axons where track disruption drives neurodegenerative disease. We develop an \(N\)-track totally asymmetric simple exclusion process (TASEP) model where motors reversibly detach into local off-track states. Using mean-field, continuum, and fast-switching approximations, we derive an effective nonlinear transport equation for total motor density. This reduced model retains the classical TASEP phase structure but reveals that increasing \(N\) shifts the boundaries between low-density, high-density, and maximal-current regimes. Crucially, under source-driven boundary conditions reflecting constant cargo production, the entrance density undergoes a saddle-node bifurcation when the production rate exceeds the network's transport capacity, \(N/4\). This triggers persistent cargo accumulation that propagates throughout the domain. Our results establish a physical mechanism by which gradual microtubule degradation drives a transition from efficient transport to jammed pathological states.
\end{abstract}

\textit{Introduction.} Intracellular active transport via molecular motors along microtubules is essential for cellular function, particularly in extended structures like neuronal axons. In neurodegenerative pathologies such as Alzheimer's, neurons exhibit deficient transport~\cite{berth,yonekawa}, leading to the toxic accumulation of cellular cargo~\cite{tanabe}. In this Letter, we examine how the number of tracks, representing the number of microtubules traversing an axon, affects macroscopic transport properties.

In this context, previous studies often utilized independent drift-diffusion models~\cite{bressloff_qss,bressloff_rev,karamched2017effects,kuznetsov2022bidirectional,ciocanel2018modeling,ciocanel2026both}. However, such models neglect local exclusion and motor crowding. Consequently, they erroneously predict an effective flux that trivially scales with $N$, failing to capture the jamming observed \textit{in vivo}~\cite{encalada2014biophysical}. 

To account for local motor interactions, we utilize the Totally Asymmetric Simple Exclusion Process (TASEP)~\cite{blythe,chou2011non,tauber2014critical}. By formulating an $N$-track TASEP coupled to a localized boundary buffer, we derive a single effective PDE that reveals how the structural capacity of the network fundamentally dictates boundary conditions and transport phases. Our model provides a framework for understanding sudden onset of intraneuronal jamming in neurodegenerative pathologies.

{\textit{Linear Model.}} Consider a population of motors each performing an independent drift-diffusion process on a 1D semi-infinite track. The macroscopic PDE description of the ensemble of motors is given by a drift-diffusion equation:
\begin{equation}
    \frac{\partial\rho}{\partial t} = -v\frac{\partial\rho}{\partial x} + D\frac{\partial^2\rho}{\partial x^2},
    \label{eq:advection-diffusion}
\end{equation}
where $\rho$ represents the motor density and $v,D$ are the drift and diffusion constants, respectively. Such a model can be derived from a detailed biophysical model, where a motor-cargo complex along a microtubule shifts between ballistic $\pm$ phases and a stationary phase according to a Markov process, yielding a stochastic hybrid system. Invoking a quasi-stationary approximation then yields an drift-diffusion equation~\cite{bressloff_qss,bressloff_rev,bressloff2017stochastic}. To generalize this framework to account for the availability of multiple lanes, we consider a system of $N \in \mathbb{N}$ tracks and let $\rho_{i}(x,t)$, $i=1,...,N$, denote the motor density on the $i$th track at time $t$ and location $x$. Each $\rho_i$ evolves according to an independent drift-diffusion process. To facilitate movement of motors between tracks, we further introduce an off-track state, $\rho_0(x,t)$ which is completely stationary (no drift or diffusion). 

To make this analytically tractable, we set $\rho_T \equiv \sum_{i=1}^N \rho_i$, which yields a two-variable system:
\begin{align}
\begin{split}
    \frac{\partial\rho_0}{\partial t} &= -N\omega_A\rho_0 +\omega_D\rho_T\\
    \frac{\partial\rho_T}{\partial t} &= -v\frac{\partial\rho_T}{\partial x} + D\frac{\partial^2\rho_T}{\partial x^2} + N\omega_A\rho_0-\omega_D\rho_T \label{eq:reduced_linear_tracks},
\end{split}
\end{align}
where $\omega_A$ and $\omega_D$ are the rates of attachment and detachment from the tracks, respectively. 

We can obtain insight into how the effective flux of motors evolving according to Eqs.~\eqref{eq:reduced_linear_tracks} scales with $N$ by assuming $\omega_A, \omega_D \gg 0$ and projecting all the dynamics as a perturbation away from the reaction equilibrium:
\begin{equation}
    \rho_0\approx\frac{\omega_D}{N\omega_A}\rho_T.
\end{equation}
Let $\rho=\rho_0+\rho_T$. Projecting Eq.~\eqref{eq:reduced_linear_tracks} to the stationary measure dictated by the reaction kinetics yields, to leading order, a single effective equation:
\begin{equation}
    \frac{\partial\rho}{\partial t} =-v\frac{N\omega_A}{N\omega_A+\omega_D}\frac{\partial\rho}{\partial x}+D\frac{N\omega_A}{N\omega_A+\omega_D}\frac{\partial^2\rho}{\partial x^2},
\end{equation}
with the flux given by
\begin{equation}
    J=v\frac{N\omega_A}{N\omega_A+\omega_D}\rho-D\frac{N\omega_A}{N\omega_A+\omega_D}\frac{\partial\rho}{\partial x}.
\end{equation} 
Thus, in a linear model, the flux increases linearly with the particle concentration, $\rho$. The effect of the number of available tracks is minimal. Once $N$ is large enough, the behavior is the same as a single track where particles never leave. This minor impact is effectively an artifact of more tracks helping particles spend more time in the drifting state. The model shows no significant change in the behavior of the system as $N$ varies, which is not in line with the biological reality that microtubule tracks are necessary for active transport to occur and efficiently deliver selected cargo to specific locales. It is for this reason that we elect to use a nonlinear TASEP framework.

\textit{Multi-Track TASEP.} In the $N$-track formulation of the TASEP \footnote{See Supplementary Material (SM) for detailed background on the basic TASEP and a discussion of how the approximations in this manuscript fare in distinct parameter regimes.}, each track will be a one dimensional lattice with sites labeled $i=1,2,...,M$. Along each track, an ensemble of particles undergoes a TASEP with one important exception: a particle may unbind from the track and join an off-track particle reservoir. 
    
    We characterize the microscopic configurations of the individual lattices with occupation numbers $n_i^j$. If the $i$-th site on track $j$ is occupied, we say $n_i^j=1$, otherwise $n_i^j=0$. Let $n_i^0$ indicate the number of particles sitting in the off-track state at lattice site $i$. The following rules dictate how particles can evolve on the lattice: (1) A particle at the $i$-th site on track $j$ ($n_i^j = 1$) can jump to space $i+1$ with rate $1$ if $n_{i+1}^j=0$; (2) A particle enters track $j$ at the first site at a rate $\alpha$ if $n_1^j=0$; (3) A particle leaves track $j$ at $M$ at a rate $\beta$ if $n_M^j=1$; (4) A particle on the track can fall off the track into a local reservoir state at a rate $\omega_D$; (5) A particle sitting off of the track can reattach to any unoccupied track at a rate $\omega_A$.
   
    The first three rules described are the same as the standard TASEP model. The last two relating to the reservoir states are added to facilitate movement between the tracks \cite{parmeggiani}. See Figure~\ref{fig:transition_diagram} for a summary diagram. 

    The equations of motion of the lattice occupation numbers are characterized by Ehrenfest equations that constitute a non-trivial, many-body problem which depends on higher-order moments~\cite{bressloff2016model,chen2017semidefinite,ogura2018second}. We invoke a mean-field approximation, a continuum limit, and an isotropic track approximation (see SM) to obtain the following nonlinear drift-diffusion equation for the motor density dynamics across $N$ tracks:
    
\begin{figure} 
    \centering
    \includegraphics[width=\linewidth]{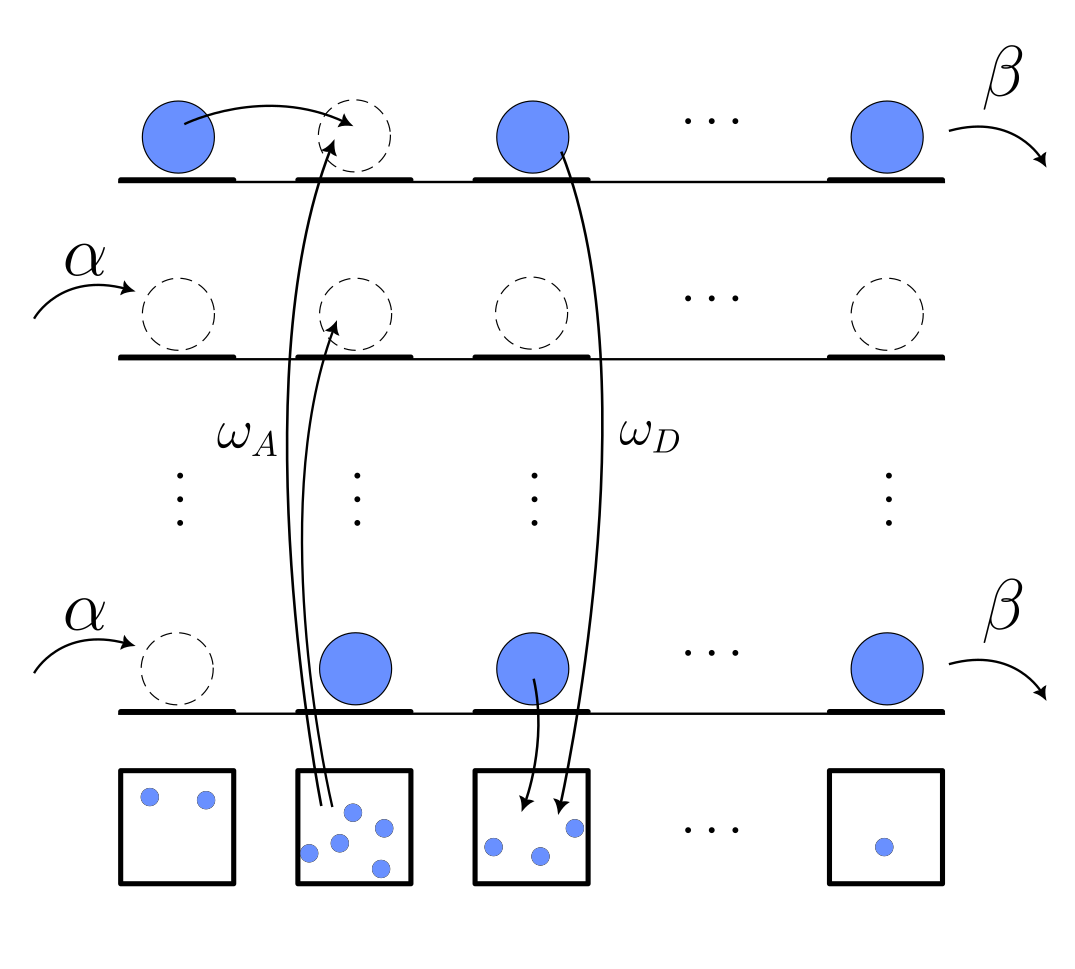}
    \caption{Transition Diagram for the $N$-Track Model}
    \label{fig:transition_diagram}
\end{figure}

\begin{subequations} \label{eq:2dsystem}
\begin{align}
     \frac{\partial\rho_0}{\partial \tau} &= -N\Omega_A\rho_0(1-\bar{\rho}) + N\Omega_D\bar{\rho} \\
     \frac{\partial\bar{\rho}}{\partial \tau} &= -(1-2\bar{\rho})\frac{\partial\bar{\rho}}{\partial x} + \frac{\varepsilon}{2}\frac{\partial^2 \bar{\rho}}{\partial x^2} +\Omega_A\rho_0(1-\bar{\rho}) -\Omega_D\bar{\rho},
\end{align}
\end{subequations}
where $\rho_0(x,t)$ represents the off-track stationary motor density, $\bar{\rho}(x,t) \equiv N^{-1} \sum_{i=1}^N \rho_i(x,t)$, $\varepsilon$ is the infinitesimal lattice spacing, $\tau = \varepsilon t$, and $\Omega_{A,D} = \varepsilon \omega_{A,D}$. Here, $\rho_i(x,t)$ denotes the motor density of motors along the $i$th track.

    Next we reduce this to a single effective equation so that we may glean some analytic insight into the dynamics. To this end, we replace the attachment and detachment rates with $\Omega_A/\delta,\Omega_D/\delta,  $ with $\delta\ll1$, rendering the transitions between on-track and off-track states to be fast. This is motivated by findings in the peripheral sensory neurons of \textit{Drosophila} that motor-cargo complexes undergo quick transitions between motile and stationary states~\cite{maeder2014vivo,wong2012neuropeptide}. We can thus invoke an adiabatic approximation by projecting the dynamics onto the stationary measure of the Markov process dictating the on-track/off-track transitions and treat the spatial dynamics as a perturbation away from this quasi-steady state. 
    
    Balancing the reaction terms yields
\begin{subequations}
    \begin{align}
        \rho_0 = \frac{\Omega_D\bar{\rho}}{\Omega_A(1-\bar{\rho})} \\
        \bar{\rho} = \frac{\Omega_A\rho_0}{\Omega_A\rho_0 + \Omega_D}.
        \label{eq:rxn_balance}
    \end{align}        
    \end{subequations}
Define $\rho(x,t)$ as the total motor concentration. At quasi-stationarity, we can express $\rho$ in terms of the on-track or off-track motor densities:
\begin{subequations} 
    \begin{align}
        \rho(x,t) &= f_0(\rho_0) = f_a(\bar{\rho}) \\
        f_0(\rho_0) &\equiv N \frac{\Omega_A\rho_0}{\Omega_A\rho_0 + \Omega_D} + \rho_0\\
        f_a(\bar{\rho}) &\equiv N\bar{\rho} + \frac{\Omega_D\bar{\rho}}{\Omega_A(1-\bar{\rho})}.
    \end{align}  
    \label{eq:total_1}
\end{subequations}
Equations~\eqref{eq:total_1} are obtained by computing $N\bar{\rho} + \rho_0$ as calculated at quasi-stationarity, Eqs.~\eqref{eq:rxn_balance}. The functions, $f_0:[0,\infty) \rightarrow [0,\infty)$, $f_a:[0,1) \rightarrow [0,\infty)$ are both invertible on their domains. Thus, we can add Eqs.~\eqref{eq:2dsystem}. This yields the conservation equation: 
    \begin{equation} \label{eq:1d_conservation}
        \frac{\partial \rho}{\partial t} = -N\frac{\partial}{\partial x}[\bar{\rho}(1-\bar{\rho})] + \frac{N\varepsilon}{2}\frac{\partial^2\bar{\rho}}{\partial x^2}
    \end{equation}
Next we perform the adiabatic approximation around the nonlinear stationary measure. Specifically, this will allow us to express the motor densities along individual tracks as a function of the total density. Define the expansion 
    \begin{align}
        \rho_j(x,t) = R_j(\rho(x,t))+\delta\mu_j(x,t)+O(\delta^2).
    \end{align}
Here, $R_0(\rho) = f_0^{-1}(\rho)$ and $R_j(\rho) = f_a^{-1}(\rho)$ for $j \geq 1$ and $\mu_j(x,t)$ characterize the fluctuations around the equilibrium with $\sum_j \mu_j = 0$.
Substituting leading order approximation $\rho_i \approx R_i(\rho)$ into the conservation equation gives us
    \begin{align} \label{eq:first_order}
    \frac{\partial \rho}{\partial t} &\approx - \frac{\partial}{\partial x} \underbrace{\left[ \sum_{i=1}^N R_i(\rho) (1 - R_i(\rho)) \right]}_{\Phi(\rho)} + \frac{\varepsilon}{2} \frac{\partial^2}{\partial x^2} \underbrace{\left[ \sum_{i=1}^N R_i(\rho) \right]}_{D_{\text{int}}(\rho)}.
    \end{align}
We perform an adiabatic elimination of the fast reaction modes by expanding $\rho_j = R_j(\rho) + \delta \mu_j$. Linearizing the reaction terms and inverting the Jacobian on the subspace orthogonal to the conservation vector yields a first-order perturbation that generates a flux correction (see SM). The resulting effective equation for the total density is:
    
    \begin{equation} \label{eq:avgflow}
        \left(1+\frac{\Omega_D/N\Omega_A}{(1-\bar{\rho})^2}\right)\frac{\partial \bar{\rho}}{\partial \tau} = -\frac{\partial}{\partial x}[\bar{\rho}(1-\bar{\rho})] + \frac{\varepsilon}{2}\frac{\partial^2\bar{\rho}}{\partial x^2}
    \end{equation}
    
Comparing Eq.~\eqref{eq:avgflow} with the hydrodynamic limit of the standard TASEP suggests that the reduced system acts like a basic TASEP model with a nonlinear timescale coefficient. This allows us to draw many parallels between the two. In particular, this means that the equilibrium profile, with $\rho_t=0$, will look identical to the classic TASEP equilibrium profile provided the boundary conditions are the same.
    
\textit{Multi-track Phase Diagram.} In the classic TASEP, the equilibirum flux is completely determined by the boundary values, $\rho(0) = \alpha$ and $\rho(L) = 1- \beta$. In $(\alpha,\beta)$ parameter space, the equilibrium motor densities and fluxes are partitioned into three regions: High Density (HD), Low Density (LD), or Maximal Current (MC). HD indicates that the density in the bulk of each track is above $1/2$, while LD indicates that it is below $1/2$. In each case the flux is less than the maximum value of $1/4$ attained in the MC phase, where the bulk density is $1/2$. 

In the multi-track case, the effective conservation equation (Eq.~\eqref{eq:1d_conservation}), is exactly the hydrodynamic limit of the standard TASEP, except the flux is multiplied by $N$. Thus, for the multi-track model, the maximal flux is $N/4$. 

To properly partition the parameter space of the multi-track model into distinct regions of equilibrium motor density and flux, we next need to characterize the boundary conditions of the multi-track conservation equation, Eq.~\eqref{eq:1d_conservation}.

\begin{figure} 
    \centering
    \includegraphics[width=1.0\linewidth]{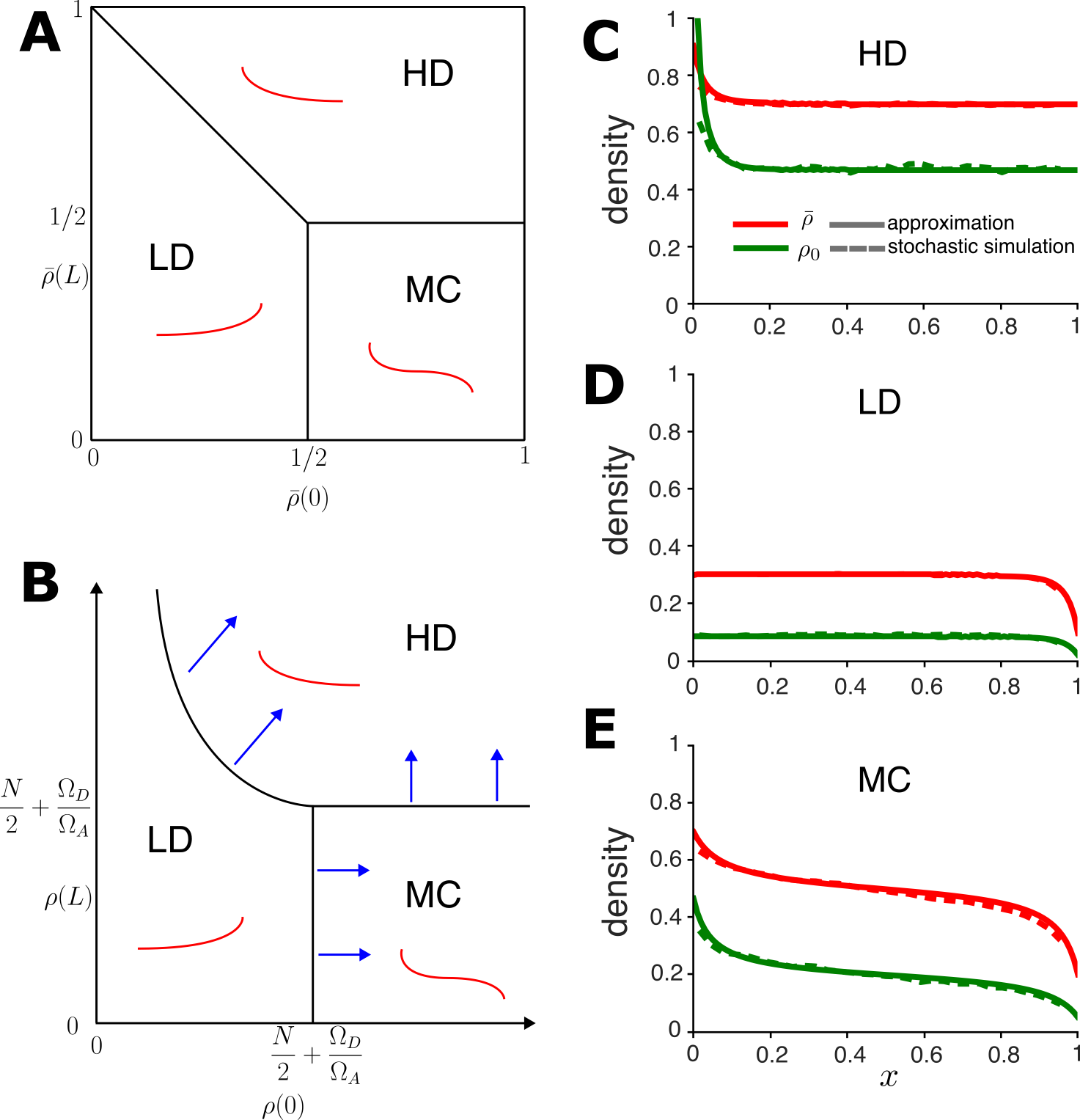}
    \caption{Multiple phases of the $N$-track TASEP. \textbf{a)} Phase of the system as a function of the track concentration at either end. \textbf{b)} Phase of the system as a function of the total concentration at either end. Red lines show the shape of the density profile in that region. Blue arrows show the movement of the boundaries as $N$ increases. \textbf{c)} High Density Phase, $\alpha=0.9,\beta=0.3$ \textbf{b)} Low Density Phase, $\alpha=0.2,\beta=0.8$ \textbf{c)} Maximal Current Phase, $\alpha=0.7,\beta=0.8$. In all three, red lines indicate the concentration on the track and the green line indicates the concentration off the track in the respective phase. Dotted lines represent simulations using the Gillespie algorithm \cite{Gillespie}. Other parameter values are: $N=5,\Omega_A= 0.5, \Omega_D=0.1, M=32$.}
    \label{fig:NDIM_PHASE_DIAGRAMS}
\end{figure}

The motor density dynamics at $x = 0$ are given by
    \begin{align}
        \varepsilon \frac{\partial \rho(0)}{\partial \tau} &\approx N\alpha(1-\bar{\rho}(0))-N\bar{\rho}(0)(1-\bar{\rho}(0)). 
         \label{eq:lbound}
    \end{align}
Flux conservation then necessitates that $\bar{\rho}(0)=\alpha$ at equilibrium. Motor density dynamics at $x = L$ are given by 
    \begin{align}
     \varepsilon\frac{\partial\rho(L)}{\partial \tau} &= - N\beta\bar{\rho} + N\bar{\rho}(1-\bar{\rho}) 
        \label{eq:rbound}.
    \end{align}
Again, imposing flux conservation yields $\bar{\rho}(L) = 1-\beta$. 

    These boundary conditions are structurally identical to the standard TASEP model. Using Eq.~\eqref{eq:avgflow}, we can thus construct the phase diagram of the multi-track model, as the per-track equilibrium dynamics on each track are identical to the original TASEP as well. In Figure~\ref{fig:NDIM_PHASE_DIAGRAMS}a, we demonstrate this mapping with the TASEP by plotting the phase diagram as a function of the boundary conditions in $(\bar{\rho}(0), \bar{\rho}(L))$ parameter space (Figure~\ref{fig:NDIM_PHASE_DIAGRAMS}a). From this, we show the phase diagram in individual on-track motor density space, $(\rho(0), \rho(L))$, by applying the mapping $f_a(\bar{\rho})$. This allows us to demonstrate exactly how $N$ modifies the geometry of the boundaries between the distinct phases. In Figure~\ref{fig:NDIM_PHASE_DIAGRAMS}b-e, we show the result. As $N$ increases, the boundaries between the distinct phases move so that the LD region grows and the HD region shrinks. The MC region requires a higher $\rho(0)$ value for larger $N$ but becomes increasingly insensitive to the motor density at $x=L$.

While these initial results provide a useful mathematical baseline, they lack biological realism because they implicitly assume an infinite reservoir of motors at the boundary. We can see this by examining the flux in each phase. The first-order approximation (Eq. (11)) can be written as $\rho_t + J_x = 0$, where the flux is $J(\rho) = N\bar{\rho}(1-\bar{\rho}) - N\frac{\epsilon}{2}\bar{\rho}_x$. At steady state ($\rho_t = 0$), $J$ is spatially constant. In the bulk of the domain, near the densities seen in Figure 2c-d, the density gradient vanishes ($\bar{\rho}_x \approx 0$), reducing the flux to simply $J = N\rho_b(1-\rho_b)$, where $\rho_b(x)$ is the bulk on-track density. Under these boundary conditions, $\alpha$ and $\beta$ strictly fix the per-track bulk density $\rho_b$, meaning $\rho_b$ is entirely independent of $N$.
Consequently, the total flux grows linearly with $N$. Physically, this implies each microtubule acts as a strictly independent pump, drawing from an infinite, private reservoir. Under such idealized conditions, the tracks never truly interact, and the shifting phase boundaries in Figure~\ref{fig:NDIM_PHASE_DIAGRAMS}b are merely a geometric artifact of mapping into total-density coordinates. But neurons do not summon cargo \textit{ex nihilo}. To capture the true physical bottleneck of axonal transport, the tracks must compete for a shared supply. In the next section, we abandon these standard open boundaries and couple the entire network to a single, localized somatic buffer. By forcing the tracks to drain a common pool, we reveal how structural limitations transform a simple queue into a traffic jam. 
    
\textit{Alternate Boundary Conditions.} To implement this localized boundary buffer constraint, we adjust the boundary conditions to reflect localized production and consumption. At the left boundary, particles are now injected at a rate $\alpha$ directly into the off-track state, $n_0^0$. This naturally utilizes the reservoir built into the model and permits particle buildup—a phenomenon that can only occur in the standard model if $\alpha > 1$. At the right boundary, we assume particles exit solely from the active track. While exiting from the reservoir or both states is theoretically possible, exiting only from the track simplifies the analysis without fundamentally altering system dynamics.

\begin{figure*}[htbp!]
    \includegraphics[width = 1\textwidth]{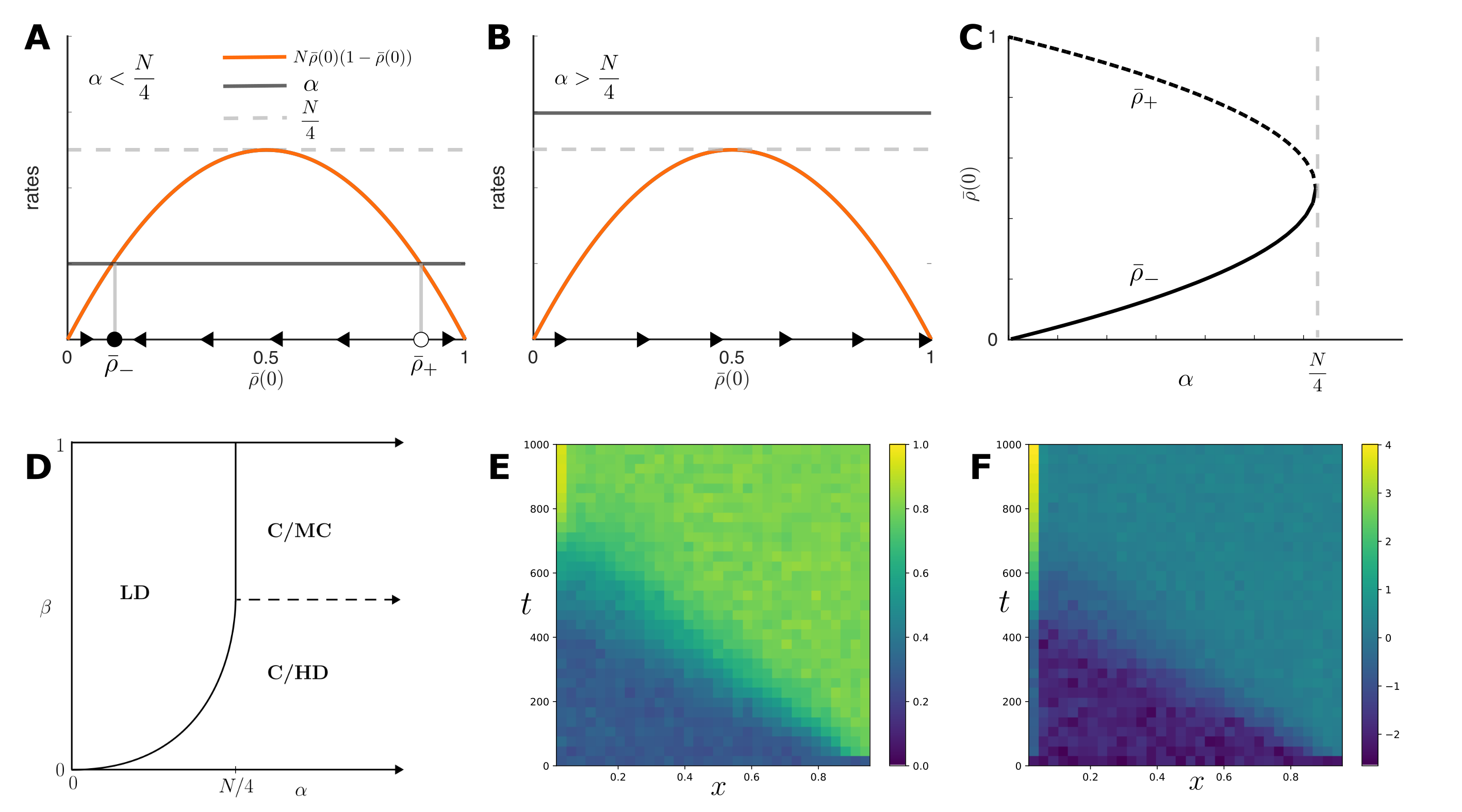}
    \caption{{A saddle-node bifurcation point determines when sudden jamming occurs. (A) For $\alpha < N/4$, two equilibria exist, $\bar{\rho}_{\pm}$, with $\bar{\rho}_-$ stable ($\Large{\bullet}$) and $\bar{\rho}_+$ unstable ($\Large{\circ}$). (B) When $\alpha > N/4$, no equilibria exist and the off-track motor reservoir ($\rho_0$) grows without bound. (C) A bifurcation diagram demonstrating that the equilibria collide and annihilate one another at exactly $\alpha = N/4$, where a saddle-node bifurcation occurs.} (D) A phase diagram in the $(\alpha,\beta)$ plane is shown. LD indicates no buildup at the left boundary. C/MC indicates buildup due to overproduction while C/HD indicates buildup due to under-consumption. (E) A heatmap showing the track density in the C/HD phase. A wave propagates from the right boundary and disrupts the stable equilibrium at the left boundary. (F) A heatmap showing the log of the off-track reservoir values in the same simulation. Unbounded buildup is seen on the left. Simulations use $N = 4, \alpha=0.8, \beta=0.2$.}
    \label{fig:sn_bif}
\end{figure*}

Because the exit dynamics are unchanged from Eq.~\eqref{eq:rbound}, the equilibrium density at the right boundary is $1-\beta$ for $0 < \beta \le 1$. For exit rates $\beta > 1$, the boundary clears instantly, clamping the density at $0$.  At the left boundary
    \begin{align}
       \varepsilon \frac{\partial \bar{\rho}(0)}{\partial \tau} &\approx \alpha-N\bar{\rho}(0)(1-\bar{\rho}(0)).
        \label{eq:lbound_alt}
    \end{align}
If $\alpha<N/4$, there are two equilibrium motor densities at $x = 0$ that satisfy $\rho_\tau=0$:

    \begin{equation} \label{eq:ss_vals}
        \bar{\rho}_{\pm}(0) = \frac{1}{2}\left(1\pm \sqrt{1-\frac{4\alpha}{N}}\right),
    \end{equation}
with $\bar{\rho}_-$ being stable and $\bar{\rho}_+$ being unstable. Thus, the model predicts unbounded buildup at $x = 0$ if the motor density were perturbed above $\bar{\rho}_+$. In the stochastic case, this will happen if the value at the right boundary is larger than $\bar{\rho}_+$. This can be seen in the expressions for wave velocity, $v_g$, and shock velocity~\cite{salsa2016partial,ockendon2003applied}, $v_s$, coming from $\eqref{eq:avgflow}$.
\begin{align}
v_g &= \frac{1 - 2\overline{\rho}}{1 + \frac{\Omega_D / N\Omega_A}{(1 - \overline{\rho})^2}} \label{eq:vg} \\
v_s &= \frac{1 - \overline{\rho}_L - \overline{\rho}_R}{1 + \frac{\Omega_D / N\Omega_A}{(1 - \overline{\rho}_L)(1 - \overline{\rho}_R)}} \label{eq:vs}
\end{align}
Here $\bar{\rho}_{L/R}$ indicates the concentration on the Left/Right side of the shock. With $\bar{\rho}_L = \bar{\rho}_-$ and $\bar{\rho}_R>\bar{\rho_+}$, we can see that a wave comes from the right side and meets the low density wave from the left. The shock moves left and pushes that boundary higher, above the critical $\rho_+$ value (Figure~\ref{fig:sn_bif}E-F). This means our stability condition is $1-\beta>\bar{\rho}_+ \iff \beta<\bar{\rho}_-$. Thus, the left boundary is in a state of unbounded buildup for 

    \begin{equation} \label{eq:beta_boundary}
        \beta<\frac{1}{2}\left(1-\sqrt{1-\frac{4\alpha}{N}}\right) \text{ if } \alpha<N/4
    \end{equation}
    At $\alpha = N/4$, the boundary dynamics undergo a saddle-node bifurcation (see Fig. ~\ref{fig:sn_bif}A-C). For strictly $\alpha > N/4$, no stable equilibria exist. Because the on-track density is strictly bounded ($\bar{\rho} \le 1$), this missing equilibrium manifests as unbounded growth of the \textit{off-track} reservoir density, $\rho_0$. In this regime, the localized somatic pool overflows regardless of the value of $\beta$, saturating the entrance to the active tracks and locking the boundary density at $\bar{\rho}(0)=1$.
    
    This gives us a new phase diagram in terms of the production rate $\alpha$ and the consumption rate $\beta$ shown in Figure \ref{fig:sn_bif}. LD indicates the familiar low density phase, and C indicates the phase where the particles get jammed at the left boundary. The HD and LD sub-phases of C  indicate the bulk density on the track away from the buildup.

    To determine how the number of tracks, $N$, impacts transport capacity, recall that the equilibrium flux is $J = N\rho_b(1-\rho_b)$, where $\rho_b$ represents the bulk on-track density. In the LD phase, the bulk density is dictated by the left boundary, meaning $\rho_b = \rho_-$. Substituting our expression for $\rho_-$ from Eq.~\eqref{eq:ss_vals} reveals that the flux is entirely source-limited, yielding $J = \alpha$. Conversely, the C phase is characterized by unbounded particle accumulation in the off-track reservoir. This persistent buildup saturates the entrance to the active tracks, locking the boundary density at $\bar{\rho}(0) = 1$. In the MC and HD sub-phases, $\rho_b=1/2$ and $1-\beta$ respectively. A brief calculation shows that in both cases $J=\alpha_{\text{crit}}$, the critical value of alpha on the boundary, which is also the highest flux achievable in the LD phase. Thus, excess production only contributes to the congestion rather than increasing the flux.

    A natural follow-up would be to consider spatial heterogeneity of number of microtubule tracks. What if one region of the axon has $N_1$ tracks and the next has $N_2$ tracks with $N_1> N_2$? If motors are flowing from $N_1$ tracks to $N_2$, jamming occurs and propagates into the more numerous microtubule region. This suggests that microtubule deficiency in any region of the axon could cause the onset of jamming (see SM for details).

    Ultimately, this localized boundary buffer exposes the tipping point of axonal transport. The network does not merely slow down as microtubules degrade; it undergoes a fundamental phase transition. If the number of available tracks, $N$, decreases due to neurodegenerative decay, the critical capacity threshold of the system, $N/4$, shrinks alongside it. A healthy neuron steadily clearing cargo in the source-limited LD phase can suddenly find its normal production rate outpacing its compromised infrastructure ($\alpha > N/4$). The mathematical consequence is the saddle-node bifurcation that forces the system into the C phase. Biologically, this manifests as a devastating traffic jam: cargo pools uncontrollably in the local reservoir, saturating the entrances of the remaining microtubules and mirroring the pathogenic protein buildup observed in Alzheimer's disease.

\textit{Conclusion.} In a theoretical light, our $N$-track model demonstrates that unveiling the macroscopic properties of multi-track systems with local particle exclusion is a physical necessity. Some previous multilane models concluded that these systems can be approximated by a single-lane TASEP with rescaled constants, but in those papers, the macroscopic properties were not dependent on a common resource pool\cite{Curatolo, Greulich,Shen,Kuan,Ezaki,Hilhorst}. By coupling the active tracks to a shared, localized boundary buffer, we show that $N$ fundamentally dictates the boundary conditions, eliciting a saddle-node transition. 

This critical transition directly addresses the mechanics of neurodegenerative decline. As pathologies like Alzheimer's disease degrade microtubules, the network's critical capacity threshold, $N/4$, shrinks. Once cellular demand outpaces this compromised infrastructure ($\alpha > N/4$), the axon suffers a transition from a healthy, source-limited phase into a maximally jammed state. The analytical solutions presented here provide the mathematical scaffolding to make further biological complexities—such as bidirectional motor populations and branching topologies—tractable, moving us closer to a comprehensive physical model of neural health.


\providecommand{\latin}[1]{#1}
\makeatletter
\providecommand{\doi}
  {\begingroup\let\do\@makeother\dospecials
  \catcode`\{=1 \catcode`\}=2 \doi@aux}
\providecommand{\doi@aux}[1]{\endgroup\texttt{#1}}
\makeatother
\providecommand*\mcitethebibliography{\thebibliography}
\csname @ifundefined\endcsname{endmcitethebibliography}
  {\let\endmcitethebibliography\endthebibliography}{}

`

\end{document}